\documentclass[12pt,a4paper]{article}
\usepackage{amsfonts}
\usepackage{amsmath}

\input{tcilatex}
\def\beq{\begin{equation}}
\def\eeq{\end{equation}}

\def\bt{\begin{tabular}}
\def\et{\end{tabular}}

\begin{document}

\author{B. Choutri, O. Cherbal, F. Z. Ighezou and M. Drir \\
{\normalsize Faculty of Physics, Theoretical Physics Laboratory,}\\
{\normalsize USTHB, B.P. 32, El Alia, Algiers 16111, Algeria.}}
\title{Pseudo-Hermitian systems with $\mathcal{PT}$-symmetry: Degeneracy and
Krein space}
\maketitle

\begin{abstract}
We show in the present paper that pseudo-Hermitian Hamiltonian systems with
even $\mathcal{PT}$-symmetry $(\mathcal{P}^{2}=1,\mathcal{T}^{2}=1)$ admit a
degeneracy structure. This kind of degeneracy is expected traditionally in
the odd $\mathcal{PT}$-symmetric systems $(\mathcal{P}^{2}=1,\mathcal{T}%
^{2}=-1)$ which is appropriate to the fermions as shown by Jones-Smith and
Mathur \cite{Smith2010} who extended $\mathcal{PT}$-symmetric quantum
mechanics to the case of odd time-reversal symmetry. We establish that the
pseudo-Hermitian Hamiltonians with even $\mathcal{PT}$-symmetry admit a
degeneracy structure if the operator $\mathcal{PT}$ anticommutes with the
metric operator $\eta $ which is necessarily indefinite. We also show that
the Krein space formulation of the Hilbert space is the convenient framework
for the implementation of unbroken $\mathcal{PT}$-symmetry. These general
results are illustrated with great details for four-level pseudo-Hermitian
Hamiltonian with even $\mathcal{PT}$-symmetry.\ 

PACS numbers: 03.65.-w, 11.30.Er

\textbf{Keywords: }Pseudo-Hermiticity, $\mathcal{PT}$-symmetry, degeneracy,
Krein space.

\ \ 
\end{abstract}

\newpage

\section{Introduction}

The research works which deal with pseudo-Hermitian and $\mathcal{PT}$%
-symmetric Hamiltonians have received a great deal of interest over the last
two decades {\normalsize \cite%
{Bender1998,Bender2002,Bender2007,Mostafa2002a,Mostafa2002b,Mostafa2002c,Mostafa2010,Croke2015,Brody2016}%
}.{\normalsize \ }In this context, Sato and al {\normalsize \cite{Sato2012}}
established a generalization of the Kramers degeneracy to pseudo-Hermitian
Hamiltonians admitting even time-reversal symmetry $(\mathcal{T}^{2}=1)$.
This extension is achieved using the mathematical structure of
split-quaternions\footnote{%
The quaternion algebra is generated by the $2\times 2$ unit matrix $\sigma
_{0}$ and the pure imaginary complex number $i$ multiplied by the SU(2)
Pauli matrices $(i\sigma _{x},i\sigma _{y},i\sigma _{z}),$\ while the
split-quaternion algebra is generated by the $2\times 2$ unit matrix $\sigma
_{0}$ and the pure imaginary complex number $i$ multiplied by the SU(1,1)
Pauli matrices $(-\sigma _{x},-\sigma _{y},i\sigma _{z}).$} instead of
quaternions, usually adopted in the case of Hermitian Hamiltonians with odd
time-reversal symmetry $(\mathcal{T}^{2}=-1)$ \cite{Avron1989}. In a recent
paper {\normalsize \cite{Cherbal2014}}, we have found that the metric
operator for the pseudo-Hermitian Hamiltonian $H$ that allows the
realization of the generalized Kramers degeneracy is necessarily indefinite.
We have further shown that such $H$ with real spectrum also possesses odd
antilinear symmetry induced from the existing odd time-reversal symmetry of
its Hermitian counterpart $h$, so that the generalized Kramers degeneracy of 
$H$ is in fact crypto-Hermitian Kramers degeneracy {\normalsize \cite%
{Cherbal2014}}. \ \ \ \ \ \ \ \ \ \ \ \ \ \ \ \ \ \ \ \ \ \ \ \ \ \ \ \ \ \
\ \ \ \ \ \ \ \ \ \ \ \ 

On the other hand, as it is well known in\emph{\ }$\mathcal{PT}$ quantum
theory {\normalsize \cite{Bender2007,Croke2015}}\emph{,} due to the
antilinearity of $\mathcal{PT}$,\emph{\ }the eigenstates of a $\mathcal{PT}$%
-symmetric Hamiltonian $H$, may or may not be the eigenstates of\emph{\ }$%
\mathcal{PT}$\emph{. }If every eigenstate of\emph{\ }$H$\emph{\ }is also an
eigenstate of $\mathcal{PT}$, then we have an unbroken\emph{\ }$\mathcal{PT}$%
-symmetry, which corresponds to real eigenvalues\emph{. }Conversely, if some
of the eigenstates of a $H$ are not simultaneously eigenstates of $\mathcal{%
PT}$\emph{, }then we have a broken $\mathcal{PT}$-symmetry, which
corresponds to complex conjugate eigenvalues. The\emph{\ }$\mathcal{PT}$%
-broken\emph{\ }and $\mathcal{PT}$-unbroken\ phases are separated by
exceptional points (EPs) which are non-Hermitian degenerate points with
coalesced eigenvalues and eigenvectors {\normalsize \cite{Heiss2004}}\emph{. 
}For $\mathcal{PT}$-symmetric systems, the exceptional points are obligatory
passage in the $\mathcal{PT}$-broken and $\mathcal{PT}$-symmetric phase
transitions. In this context, Berry has shown {\normalsize \cite{Berry2004} }%
that non-Hermitian physics differs radically from Hermitian physics in the
presence of degeneracies. This non-Hermitian behavior has been further
illustrated in several physical examples{\normalsize \ \cite{Berry2004}}.
Recently, the role of degeneracy in $\mathcal{PT}$-symmetry breaking has
been also subject of investigations {\normalsize \cite{Ge2014}.}\ \emph{\ }%
The degeneracy for odd $\mathcal{PT}$-symmetric systems $(\mathcal{P}^{2}=1,%
\mathcal{T}^{2}=-1)$ appropriate to the fermions has been studied \cite%
{Smith2010}. It has been established \cite{Smith2010} that an analog of
Kramers degeneracy exists also for odd $\mathcal{PT}$-symmetric systems, and
an unbroken $\mathcal{PT}$-symmetry can exist if we assemble the two column
vectors of the $\mathcal{PT}$ doublets in a single quaternionic column.

The purpose of the present paper is to extend the non-Hermitian degeneracy
behavior developed for odd $\mathcal{PT}$-symmetric systems \cite{Smith2010}
to even{\normalsize \ }$\mathcal{PT}${\normalsize -}symmetric ones $(%
\mathcal{P}^{2}=1,\mathcal{T}^{2}=1)$. \ \ \ \ \ \ \ \ \ \ \ \ \ \ \ \ \ \ \
\ \ \ \ \ 

The paper is organized as follows.\thinspace\ In section 2, we first analyze
the existence of the degeneracy structure for pseudo-Hermitian Hamiltonian
systems with even $\mathcal{PT}$-symmetry. We{\normalsize \ }will establish
that the degeneracy structure exists if the $\mathcal{PT}$\ operator
anticommutes with the metric operator $\eta $, which is necessarily
indefinite metric. Moreover, we emphasize in section 3 the role of the Krein
space formulation of the Hilbert space for restoring an unbroken $\mathcal{PT%
}$-symmetry. This degeneracy structure is thus well implemented in the Krein
space formulation of the Hilbert space. In the purpose of illustration of
the above general results, we study in great details in section 3, the
pseudo-Hermitian four-level Hamiltonian invariant under the even $\mathcal{PT%
}$-symmetry.\ The paper ends with conclusion and outlook. \ \ \ \ \ \ \ \ \
\ \ \ \ \ 

\section{Pseudo-Hermiticity, $\mathcal{PT}$-symmetry and degeneracy}

We start our analysis by considering a diagonalizable pseudo-Hermitian
Hamiltonian $H$ with finite-dimensional spectrum. A complete biorthonormal
eigenbasis $|\psi _{n}\rangle ,|\phi _{n}\rangle $ exists, i.e. a basis such
that {\normalsize \cite{Mostafa2002a,Mostafa2010, Wong, Faisal} }\ \ \ \ \ \
\ \ \ \ 
\begin{eqnarray}
H\left\vert \psi _{n}\right\rangle &=&E_{n}\left\vert \psi _{n}\right\rangle
,\text{ \ }H^{\dagger }\left\vert \phi _{n}\right\rangle =E_{n}^{\ast
}\left\vert \phi _{n}\right\rangle ,  \label{021} \\
\langle \psi _{n}\left\vert \phi _{m}\right\rangle &=&\langle \phi
_{n}\left\vert \psi _{m}\right\rangle =\delta _{nm},  \label{022} \\
\underset{n}{\sum }\left\vert \phi _{n}\right\rangle \langle \psi _{n}|\text{
} &=&\underset{n}{\sum }\left\vert \psi _{n}\right\rangle \langle \phi _{n}|%
\text{ }=\mathbf{1}.\text{ \ }  \label{023}
\end{eqnarray}%
Here we deal with real eigenvalues of $H.$ This yields the following
spectral representation of $H$ and the metric operator $\eta $ {\normalsize 
\cite{Mostafa2002a,Mostafa2010}}: \ \ 
\begin{equation}
H=\underset{n}{\sum }E_{n}\left\vert \psi _{n}\right\rangle \langle \phi
_{n}|,\text{ \ \ \ \ }\eta =\underset{n}{\sum }\left\vert \phi
_{n}\right\rangle \langle \phi _{n}|,  \label{024b}
\end{equation}%
with the properties 
\begin{equation}
\left\vert \phi _{n}\right\rangle =\eta \left\vert \psi _{n}\right\rangle ,%
\text{ }|\psi _{n}\rangle =\eta ^{-1}\left\vert \phi _{n}\right\rangle .
\label{025}
\end{equation}%
Since $H$ is pseudo-Hermitian, this means that $H$ satisfies the relation $%
H^{\dagger }=\eta H\eta ^{-1}$, where $\eta $ is an Hermitian bounded linear
invertible operator with bounded inverse operator. Furthermore, we assume
that $H$ is$\,$invariant under the even $\mathcal{PT}$-symmetry, i.e $\left[
H,\mathcal{PT}\right] =0,$ $(\mathcal{P}^{2}=1,\mathcal{T}^{2}=1)$, where $%
\mathcal{P}$ and $\mathcal{T}$ are parity and even time reversal operators
respectively. Here we define the parity operator $\mathcal{P}$ by following
the definition given in \cite{Smith2010}. The action of $\mathcal{P}$ to any
state $\psi $ is to multiply $\psi $ by the matrix $S$ which is a real $N$%
-dimensional matrix given by \cite{Smith2010}, 
\begin{equation}
S=\left( 
\begin{array}{cc}
I & 0 \\ 
0 & -I%
\end{array}%
\right) \text{,}
\end{equation}%
where $I$ denotes the $(N/2)$-dimensional identity matrix. For the
time-reversal operator, one follows the definition given in {\normalsize 
\cite{Sato2012}} as $\mathcal{T}=ZK\mathit{,}$ with $K$ being a complex
conjugation operator, $Z$ is an unitary matrix which can be chosen as a real
matrix with all the diagonal terms equal to a $2\times 2$ Pauli matrix $%
\sigma _{x}$ and all the off-diagonal terms equal to zero {\normalsize \cite%
{Sato2012}}, namely \ \ \ \ 
\begin{equation}
Z=\left( 
\begin{array}{cccc}
\sigma _{x} &  &  &  \\ 
& . &  &  \\ 
&  & . &  \\ 
&  &  & \sigma _{x}%
\end{array}%
\right) \text{, \ \ }\sigma _{x}=\left( 
\begin{array}{cc}
0 & 1 \\ 
1 & 0%
\end{array}%
\right) \text{.}
\end{equation}%
We have $\mathcal{T}^{2}=1$. It is useful to note that as established \cite%
{Smith2010}, the non-Hermitian Hamiltonians which are invariant under the
odd $\mathcal{PT}$-symmetry, admit Kramers degeneracy implemented by the
mathematical structure of quaternions \cite{Avron1989}. This degeneracy is
expected because the time-reversal operator is odd ($\mathcal{T}^{2}=-1$). \
\ \ \ \ \ \ 

We propose to show in the following that a degeneracy exists also for
pseudo-Hermitian Hamiltonians with even $\mathcal{PT}$-symmetry. In order to
show the degeneracy in the eigenvalues of $H$, we show that the eigenstates $%
|\psi _{n}\rangle $ and $\mathcal{PT}|\psi _{n}\rangle $ which correspond to
the same eigenvalue $E_{n}$ are linearly independent. As shown in \cite%
{Sakurai1994}, any antiunitary operator $\theta $ can be written as $\theta
=UK,$ where $U$ is an unitary operator and $K$ is the conjugation operator.
Here we have 
\begin{equation}
\theta =\mathcal{PT}=SZK=UK,
\end{equation}%
where $U=SZ$ is the unitary operator. Moreover, according to {\normalsize 
\cite{Sato2012}, }one deduce from the antiunitary property\footnote{%
The antiunitary property means that for any operator antiunitary denoted $%
\theta $, we have $\langle \theta x\left\vert \theta y\right\rangle =\langle
y\left\vert x\right\rangle $%
\par
{}} of $\mathcal{PT}$, that{\normalsize \ }$\eta $-pseudo-Hermiticity is
consistent with the $\mathcal{PT}$-symmetry if the metric operator $\eta $
commutes or anticommutes with the $\mathcal{PT}$ operator. Thus, as in 
{\normalsize \cite{Sato2012}, }the{\normalsize \ }metric operators are
classified into two: The first which commutes with $\mathcal{PT}$  and the
second which anticommutes with $\mathcal{PT}$.{\normalsize \ \ \ } \ \ \  \
\ 

Now we compute $\langle \phi _{n}\left\vert \mathcal{PT}\psi
_{n}\right\rangle ,$ where $|\psi _{n}\rangle $ and $|\phi _{n}\rangle $
form a complete biorthonormal eigenbasis (\ref{021})-(\ref{023}). By using
the antiunitary property of $\mathcal{PT}$, with the fact that $(\mathcal{PT}%
)^{2}=1$ and the relations between $|\psi _{n}\rangle $ and $|\phi
_{n}\rangle $ given in (\ref{025}), we have \ \ 
\begin{equation}
\langle \phi _{n}\left\vert \mathcal{PT}\psi _{n}\right\rangle =\langle (%
\mathcal{PT})^{2}\psi _{n}\left\vert \mathcal{PT}\phi _{n}\right\rangle
=\langle \psi _{n}\left\vert \mathcal{PT}\phi _{n}\right\rangle =\langle
\phi _{n}\eta ^{-1}\left\vert \mathcal{PT}\eta \psi _{n}\right\rangle ,
\end{equation}%
If $\mathcal{PT}$\ and $\eta $\ anticommutes, one finds from the last
equation that \ 
\begin{equation}
\langle \phi _{n}\left\vert \mathcal{PT}\psi _{n}\right\rangle =-\langle
\phi _{n}\eta ^{-1}\left\vert \eta \mathcal{PT}\psi _{n}\right\rangle .
\end{equation}%
The Hermiticity of $\eta $ leads to%
\begin{equation}
\langle \phi _{n}\left\vert \mathcal{PT}\psi _{n}\right\rangle =-\langle
\phi _{n}\left\vert \mathcal{PT}\psi _{n}\right\rangle .
\end{equation}%
We have thereby $\langle \phi _{n}\left\vert \mathcal{PT}\psi
_{n}\right\rangle =0$. On the other hand, from eq. (\ref{022}) we have $%
\langle \phi _{n}\left\vert \psi _{n}\right\rangle =1,$ so, one deduces that 
$|\psi _{n}\rangle $ and\ $\mathcal{PT}|\psi _{n}\rangle $ are linearly
independent. Consequently, as in the odd $\mathcal{PT}$-symmetric case \cite%
{Smith2010}, we have also two fold degeneracy in the eigenstates of $H$
although the $\mathcal{PT}$-symmetry is even. Here we point out that the $%
\mathcal{PT}$\ doublet $|\psi _{n}\rangle $ and \ $\mathcal{PT}|\psi
_{n}\rangle $ are linearly independent but they are not orthogonal as in the
odd $\mathcal{PT}$-symmetric case. Now we show that in the above described
scheme, the metric operator $\eta $ is necessarily indefinite. In this aim,
we shall show that the $\eta $-norm related to the $\eta $-inner product $%
\langle .\left\vert .\right\rangle _{\eta }=\langle .\left\vert \eta
.\right\rangle $ of the eigenstates of $H$ is indefinite {\normalsize \cite%
{Mostafa2003a,Sudarshan,Pauli, Lee}}. Thus we compute the norms $\langle
\psi _{n}\left\vert \psi _{n}\right\rangle _{\eta }$ and $\langle \mathcal{PT%
}\psi _{n}\left\vert \mathcal{PT}\psi _{n}\right\rangle _{\eta }$. We have \
\ \ \ \ \ 
\begin{equation}
\langle \psi _{n}\left\vert \psi _{n}\right\rangle _{\eta }=\langle \psi
_{n}\left\vert \eta \psi _{n}\right\rangle =\langle \psi _{n}\left\vert \phi
_{n}\right\rangle =1,
\end{equation}%
and \ \ 
\begin{equation}
\langle \mathcal{PT}\psi _{n}\left\vert \mathcal{PT}\psi _{n}\right\rangle
_{\eta }=\langle \mathcal{PT}\psi _{n}\left\vert \eta \mathcal{PT}\psi
_{n}\right\rangle .
\end{equation}%
By using the anticommutation relation between $\eta $ and $\mathcal{PT}$,
the antiunitary property of $\mathcal{PT}$ and $(\mathcal{PT})^{2}=1,$ one
obtains from the last equation that \ \ \ \ \ \ \ 
\begin{eqnarray}
\langle \mathcal{PT}\psi _{n}\left\vert \mathcal{PT}\psi _{n}\right\rangle
_{\eta } &=&-\langle \mathcal{PT}\psi _{n}\left\vert \mathcal{PT}\eta \psi
_{n}\right\rangle =-\langle (\mathcal{PT})^{2}\eta \psi _{n}\left\vert (%
\mathcal{PT})^{2}\psi _{n}\right\rangle   \notag \\
&=&-\langle \eta \psi _{n}\left\vert \psi _{n}\right\rangle =-\langle \psi
_{n}\left\vert \eta \psi _{n}\right\rangle =-\langle \psi _{n}\left\vert
\psi _{n}\right\rangle _{\eta }=-1.
\end{eqnarray}%
Thereby, the norm $\langle \psi _{n}\left\vert \psi _{n}\right\rangle _{\eta
}$ is positive while the norm of the partner $\langle \mathcal{PT}\psi
_{n}\left\vert \mathcal{PT}\psi _{n}\right\rangle _{\eta }$ is negative.
This means that the metric operator $\eta $ is indefinite. Due to this
degeneracy behavior, a question arizes: Is the $\mathcal{PT}$-symmetry in
the broken or unbroken phase? Traditionally, the\emph{\ }$\mathcal{PT}$%
-broken\emph{\ }and $\mathcal{PT}$-unbroken\ phases are separated by
exceptional points which are non-Hermitian degenerate points with coalesced
eigenvalues and eigenvectors. In our case, we deal with an other kind of
degeneracy different from the exceptional points degeneracy.

\textbf{Remark: }It is useful to point out that as we deal with
pseudo-Hermitian Hamiltonian $H$ with real eigenvalues, for such $H$ there
exists also a positive definite metric operator in the form of eq. (\ref%
{024b}). The latter correspond to the commutative case between the metric
operator and $\mathcal{PT}$ and does not allow therefore a degeneracy
structure. \ \ \ 

In conclusion, our Hamiltonian $H$ admit simultaneously two kinds of metric
operators, namely, a positive definite metric operator in the form of eq. (%
\ref{024b}) which does not produce a degeneracy structure, and a second
metric operator which is indefinite and achieve a degeneracy structure. \ \ 
\ \ \ \ \ \ 

\section{Krein space formulation\ }

Let us show how we achieve an unbroken $\mathcal{PT}$-symmetry. The idea is
the passage to the Krein space formulation of the Hilbert space. We recall
that in the case of even $\mathcal{PT}$-symmetry, $\mathcal{PT}$ is said to
be unbroken if the states $\left\vert \psi _{n}\right\rangle $ are invariant
under $\mathcal{PT}$, i.e $\mathcal{PT}\left\vert \psi _{n}\right\rangle
=\left\vert \psi _{n}\right\rangle $. However, the situation is different
here, although the $\mathcal{PT}$-symmetry is even, the eigenstates $%
\left\vert \psi _{n}\right\rangle $ are not invariant under $\mathcal{PT}$,
i.e $\mathcal{PT}\left\vert \psi _{n}\right\rangle \neq \left\vert \psi
_{n}\right\rangle $ because we have established previously that the $%
\mathcal{PT}$ doublets $|\psi _{n}\rangle $ and\ $\mathcal{PT}|\psi
_{n}\rangle $ are linearly independent. Then the $\mathcal{PT}$-symmetry is
broken in the Hilbert space spanned by the eigenstates of $H$. The way out
from this dead end is in fact to introduce the Krein space formulation of
the Hilbert space {\normalsize \cite{Mostafa2006, Azizov1989} }which allows
us to assemble the $\mathcal{PT}$ doublets $|\psi _{n}\rangle $ and \ $%
\mathcal{PT}|\psi _{n}\rangle $ in one single eigenstate. Indeed,%
{\normalsize \ }let us define new eigenstate $|\chi _{n}\rangle $ spanned by
the $\mathcal{PT}$ doublets $|\psi _{n}\rangle $ and \ $\mathcal{PT}|\psi
_{n}\rangle $ as follows, \ \ \ \ 
\begin{equation}
|\chi _{n}\rangle =|\psi _{n}\rangle +\mathcal{PT}|\psi _{n}\rangle ,
\end{equation}%
with the following properties, (i) the $|\chi _{n}\rangle $\ are eigenstates
of $H$ with the same eigenvalue $E_{n},$ (ii) the states $|\chi _{n}\rangle $
are invariant under $\mathcal{PT}$, i.e 
\begin{equation}
\mathcal{PT}|\chi _{n}\rangle =\mathcal{PT}(|\psi _{n}\rangle +\mathcal{PT}%
|\psi _{n}\rangle )=|\chi _{n}\rangle ,
\end{equation}%
we use the involution property of $\mathcal{PT}$, $(\mathcal{PT})^{2}=1$.
Now we show that the Hilbert space $\mathcal{K}$ spanned by\ the states $%
|\chi _{n}\rangle $ is a Krein space {\normalsize \cite{Mostafa2006,
Azizov1989}}. $\mathcal{K}$\ possess the following properties: (i) $\mathcal{%
K}$\ is endowed with the indefinite inner product $\langle .\left\vert
.\right\rangle _{\eta }$ defined in section 2. (ii) $\mathcal{K}$\ can be
decomposed in a pair of vector subspaces $\mathcal{H}_{\pm }$, where $%
\mathcal{H}_{+}$ and $\mathcal{H}_{-}$\ are spanned by the $\mathcal{PT}$
doublets $|\psi _{n}\rangle $ and \ $\mathcal{PT}|\psi _{n}\rangle $
respectively; $\mathcal{K}=\mathcal{H}_{+}\oplus \mathcal{H}_{-},$ where $%
\oplus $ means direct sum which means that for all element $f$ $\in $ $%
\mathcal{K}$ there are unique $f_{\pm }$ $\in $ $\mathcal{H}_{\pm }$ such
that $f$ $=f_{+}+f_{-}$. (iii) In the purpose of showing that $\mathcal{H}%
_{+}$ and $\mathcal{H}_{-}$ are orthogonal, we calculate the inner product $%
\langle \psi _{n}\left\vert \mathcal{PT}\psi _{n}\right\rangle _{\eta }$ for
any elements $|\psi _{n}\rangle $ $\in $ $\mathcal{H}_{+}$ and $\mathcal{PT}%
|\psi _{n}\rangle \in \mathcal{H}_{-}$, we have 
\begin{eqnarray}
\langle \psi _{n}\left\vert \mathcal{PT}\psi _{n}\right\rangle _{\eta }
&=&\langle \psi _{n}\left\vert \eta \mathcal{PT}\psi _{n}\right\rangle
=-\langle \psi _{n}\left\vert \mathcal{PT}\eta \psi _{n}\right\rangle
=-\langle (\mathcal{PT})^{2}\eta \psi _{n}\left\vert \mathcal{PT}\psi
_{n}\right\rangle  \notag \\
&=&-\langle \eta \psi _{n}\left\vert \mathcal{PT}\psi _{n}\right\rangle
=-\langle \psi _{n}\left\vert \eta \mathcal{PT}\psi _{n}\right\rangle
=-\langle \psi _{n}\left\vert \mathcal{PT}\psi _{n}\right\rangle _{\eta },
\end{eqnarray}%
thereby $\langle \psi _{n}\left\vert \mathcal{PT}\psi _{n}\right\rangle
_{\eta }=0$. This means that $\mathcal{H}_{+}$ and $\mathcal{H}_{-}$ are
orthogonal. Here, we have again used the anticommutation relation between $%
\eta $ and $\mathcal{PT}$ and the antiunitary property of $\mathcal{PT}$ and 
$(\mathcal{PT})^{2}=1$. The Hilbert space $\mathcal{K}$ is therefore a Krein
space.\ It should be noticed that the condition of unbroken $\mathcal{PT}$%
-symmetry in the odd case is different to our even $\mathcal{PT}$-symmetry
case. Indeed, in the odd case the $\mathcal{PT}$ doublets are assembled in
two column vectors $(|\psi _{n}\rangle ,$ $\mathcal{PT}|\psi _{n}\rangle )$
which form a single component column of quaternions \cite{Smith2010}. It is
not possible to assemble our $\mathcal{PT}$ doublets in two column vectors,
because the single component column obtained is a split-quaternion which
possesses a structure different from the quaternion \cite{Sato2012}.

\section{Illustration\ \ \ \ \ \ \ \ \ \ \ \ \ \ }

In this section we illustrate the above general results with an example. We
consider the four-level model described by the following non-Hermitian
Hamiltonian: \ \ 
\begin{equation}
H=\left( 
\begin{array}{cc}
a & ib \\ 
ic & -a%
\end{array}%
\right) \,,  \label{H1}
\end{equation}%
where $b=b_{0}\sigma _{0}-b_{1}\sigma _{x}-b_{2}\sigma _{y}+ib_{3}\sigma
_{z} $ and $c=b_{0}\sigma _{0}+b_{1}\sigma _{x}+b_{2}\sigma
_{y}-ib_{3}\sigma _{z} $ are real split-quaternions\footnote{%
The split-quaternion algebra is generated by the $2\times 2$ unit matrix $%
\sigma _{0}$ and the pure imaginary complex number $i$ multiplied by the
SU(1,1) Pauli matrices $(-\sigma _{x},-\sigma _{y},i\sigma _{z}).$}, and $%
a=a_{0}\sigma _{0}$ is the real split-quaternion proportional to the
identity, the $\sigma _{k}$ ($k=x,y,z$) are the Pauli matrices. By setting $%
A=b_{1}+ib_{2}$ and $B=b_{0}+ib_{3},$ this Hamiltonian $H$ can also be
written as a four-level Hamiltonian as follows: \ \ \ 
\begin{equation}
H=\,\left( 
\begin{array}{cccc}
a_{0} & 0 & iB^{\ast } & iA^{\ast } \\ 
0 & a_{0} & iA & iB \\ 
iB & -iA^{\ast } & -a_{0} & 0 \\ 
-iA & iB^{\ast } & 0 & -a_{0}%
\end{array}%
\right) ,  \label{321}
\end{equation}%
where $A^{\ast }$ and $B^{\ast }$ are complex conjugates of $A$ and $B$
respectively. The Hamiltonian $H$, in eq. (\ref{321}), satisfies the
following properties: (i) $H$ is pseudo-Hermitian, this means that $H$
satisfies the relation $H^{\dagger }=\eta H\eta ^{-1}$, (ii) $H$ is$\,$%
invariant under the even $\mathcal{PT}$-symmetry, i.e $\left[ H,\mathcal{PT}%
\right] =0,$ with $\mathcal{P}^{2}=1,\mathcal{T}^{2}=1$, where $\mathcal{P}$
and $\mathcal{T}$ are given in the case of the four-level system by \cite%
{Smith2010,Cherbal2014}, 
\begin{equation}
\mathcal{P}=\left( 
\begin{array}{cc}
I_{2} & 0 \\ 
0 & -I_{2}%
\end{array}%
\right) ,\text{ \ }\mathcal{T}=\left( 
\begin{array}{cc}
\sigma _{x} & 0 \\ 
0 & \sigma _{x}%
\end{array}%
\right) K\,,
\end{equation}%
where $I_{2}$ is the $2\times 2$ identity matrix, $\sigma _{x}$\ is Pauli
matrix, $K$ being the complex conjugation operator. Moreover, $H$\emph{\ }%
admits an indefinite metric operator $\eta $ which\ anticommutes with $%
\mathcal{PT}$, $\eta $\ is given explicitly by, \ \ \ \ \ 
\begin{equation}
\eta =\left( 
\begin{array}{cc}
\sigma _{z} & 0 \\ 
0 & -\sigma _{z}%
\end{array}%
\right) .  \label{C8}
\end{equation}%
T{\normalsize he } eigenvalues of $H$ are 
\begin{equation*}
E_{\pm }=\pm \Omega ,\quad \quad \mathrm{with}\quad \Omega =\sqrt{%
a_{0}^{2}+\left\vert A\right\vert ^{2}-\left\vert B\right\vert ^{2}}.
\end{equation*}%
This eigenvalues are indeed twofold degenerate. The Hamiltonian $H$
represents a new class of \emph{even} $\mathcal{PT}$\emph{-symmetric
Hamiltonians with degeneracy}. We remark that $H$ is asymmetric. \noindent
We deal with real eigenvalues, i.e. \ 
\begin{equation}
a_{0}^{2}+\left\vert A\right\vert ^{2}>\left\vert B\right\vert ^{2}.
\label{real Om}
\end{equation}%
The $\mathcal{PT}$ doublets $(\left\vert \psi _{-+}\right\rangle ,\left\vert
\psi _{--}\right\rangle )$ and $(\left\vert \psi _{++}\right\rangle
,\left\vert \psi _{+-}\right\rangle )$ associated to the negative and
positive eigenvalues respectively are given as follows: \ \ \ \ \ \ 

For the negative eigenvalue $E_{-}=$ $-\Omega $:$\allowbreak $%
\begin{equation}
\left\vert \psi _{-+}\right\rangle =k\left( 
\begin{array}{c}
iA^{\ast } \\ 
iB \\ 
0 \\ 
-(\Omega +a_{0})%
\end{array}%
\right) ,\text{ \ }\left\vert \psi _{--}\right\rangle =\mathcal{PT}%
\left\vert \psi _{-+}\right\rangle =k\left( 
\begin{array}{c}
-iB^{\ast } \\ 
-iA \\ 
(\Omega +a_{0}) \\ 
0%
\end{array}%
\right) ,
\end{equation}%
For the positive eigenvalue $E_{+}=$ $\Omega $: 
\begin{equation}
\left\vert \psi _{++}\right\rangle =k\,\,\left( 
\begin{array}{c}
(\Omega +a_{0}) \\ 
0 \\ 
iB \\ 
-iA%
\end{array}%
\right) ,\quad \text{ }\left\vert \psi _{+-}\right\rangle =\mathcal{PT}%
\left\vert \psi _{++}\right\rangle =k\left( 
\begin{array}{c}
0 \\ 
\Omega +a_{0} \\ 
-iA^{\ast } \\ 
iB^{\ast }%
\end{array}%
\right) .  \label{015}
\end{equation}%
The eigenstates $(\left\vert \phi _{-+}\right\rangle ,\left\vert \phi
_{--}\right\rangle )$ and $(\left\vert \phi _{++}\right\rangle ,\left\vert
\phi _{+-}\right\rangle )$ associated to $H^{\dagger }$ are obtained by the
action of $\eta $ given in eq. (\ref{C8}) to the eigenstates of $H$, they
are given for the negative eigenvalue $E_{-}=$ $-\Omega $ by: 
\begin{equation}
\left\vert \phi _{-+}\right\rangle =k\left( 
\begin{array}{c}
iA^{\ast } \\ 
-iB \\ 
0 \\ 
-(\Omega +a_{0})%
\end{array}%
\right) ,\text{ \ }\left\vert \phi _{--}\right\rangle =k\left( 
\begin{array}{c}
-iB^{\ast } \\ 
iA \\ 
-(\Omega +a_{0}) \\ 
0%
\end{array}%
\right) ,
\end{equation}%
and for the positive eigenvalue $E_{+}=$ $\Omega $ by:%
\begin{equation}
\left\vert \phi _{++}\right\rangle =k\left( 
\begin{array}{c}
\Omega +a_{0} \\ 
0 \\ 
-iB \\ 
-iA%
\end{array}%
\right) ,\quad \text{ }\left\vert \phi _{+-}\right\rangle =\,k\left( 
\begin{array}{c}
0 \\ 
-(\Omega +a_{0}) \\ 
iA^{\ast } \\ 
iB^{\ast }%
\end{array}%
\right) ,
\end{equation}%
where $k$ is the normalization constant fixed by the requirement that: 
\begin{equation}
2\Omega (\ \Omega +a_{0})\left\vert k\right\vert ^{2}=1.
\end{equation}%
These states satisfy the abnormal relations which is a consequence of the
indefinite metric $\eta $ given in (\ref{C8}), \ \ 
\begin{equation}
\langle \psi _{_{++}}\left\vert \phi _{++}\right\rangle =1,\langle \psi
_{_{+-}}\left\vert \phi _{+-}\right\rangle =-1,\text{ }\langle \psi
_{_{-+}}\left\vert \phi _{-+}\right\rangle =1,\langle \psi
_{_{--}}\left\vert \phi _{--}\right\rangle =-1,  \notag
\end{equation}%
\begin{equation}
\langle \psi _{m\alpha }\left\vert \phi _{\overline{m}\text{ }\alpha
}\right\rangle =0,\text{ }\langle \psi _{m\alpha }\left\vert \phi _{%
\overline{m}\text{ }\overline{\alpha }}\right\rangle =0,\text{ }\langle \psi
_{m\alpha }\left\vert \phi _{m\text{ }\overline{\alpha }}\right\rangle =0.
\end{equation}%
where $\alpha =\pm ,$ $m=\pm $ , $\overline{m}$ and $\overline{\alpha }$ are
the opposite signs of $m$ and $\alpha $ respectively$.$ These states satisfy
also the relations \ 
\begin{equation}
|\psi _{_{++}}\rangle \langle \phi _{_{++}}|-|\psi _{_{+-}}\rangle \langle
\phi _{+-}|+|\psi _{-+}\rangle \langle \phi _{-+}|-|\psi _{_{--}}\rangle
\langle \phi _{_{_{--}}}|=\mathbf{1.}
\end{equation}%
The (indefinite) $\eta $-norms of the $\mathcal{PT}$ doublets are given by \
\ \ \ 
\begin{equation}
\langle \psi _{++}|\psi _{_{++}}\rangle _{\eta }=\langle \psi _{++}|\phi
_{++}\rangle =1,\quad \langle \psi _{+-}|\psi _{+-}\rangle _{\eta }=\langle
\psi _{+-}|\phi _{+-}\rangle =-1,
\end{equation}%
\begin{equation}
\langle \psi _{_{-+}}\left\vert \psi _{_{-+}}\right\rangle _{\eta }=\langle
\psi _{_{-+}}\left\vert \phi _{_{-+}}\right\rangle =1,\quad \langle \psi
_{_{--}}\left\vert \psi _{_{--}}\right\rangle _{\eta }=\langle \psi
_{_{--}}\left\vert \phi _{_{--}}\right\rangle =-1.
\end{equation}%
We see that the eigenstates $\left\vert \psi _{n}\right\rangle $ are not
invariant under $\mathcal{PT}$, i.e $\mathcal{PT}\left\vert \psi
_{n}\right\rangle \neq \left\vert \psi _{n}\right\rangle ,$ the $\mathcal{PT}
$-symmetry is therefore broken. In order to restore an unbroken $\mathcal{PT}
$-symmetry,\ one introduces the Krein space formulation of the Hilbert
space. The Krein space $\mathcal{K}$ is spanned by\ the states $|\chi _{\pm
}\rangle ,$ where $|\chi _{-}\rangle $\ and $|\chi _{+}\rangle $ are linear
combination of the The $\mathcal{PT}$ doublets $(\left\vert \psi
_{-+}\right\rangle ,\mathcal{PT}||\psi _{_{-+}}\rangle )$ and $(\left\vert
\psi _{++}\right\rangle ,\mathcal{PT}\left\vert \psi _{++}\right\rangle )$
associated to the negative and positive eigenvalue respectively. Thus, in
the Krein space $\mathcal{K},$ the eigenstates are given by,

For the negative eigenvalue $E=$ $-\Omega $: 
\begin{equation}
|\chi _{-}\rangle =|\psi _{_{-+}}\rangle +\mathcal{PT}||\psi _{_{-+}}\rangle
,
\end{equation}%
for the positive eigenvalue $E_{+}=$ $\Omega $:%
\begin{equation}
|\chi _{+}\rangle =\left\vert \psi _{++}\right\rangle +\mathcal{PT}%
\left\vert \psi _{++}\right\rangle .
\end{equation}%
We see that states $|\chi _{\pm }\rangle $ are invariant under $\mathcal{PT}$%
, i.e 
\begin{equation}
\mathcal{PT}|\chi _{n}\rangle =\mathcal{PT}(|\psi _{n}\rangle +\mathcal{PT}%
|\psi _{n}\rangle )=|\chi _{n}\rangle .
\end{equation}%
One has therefore achieved an unbroken $\mathcal{PT}$-symmetry in the Krein
space $\mathcal{K}$. \ \ \ \ \ \ \ \ \ \ \ \ \ \ \ \ \ \ \ 

\section{Conclusion and outlook\ }

We have established in the present paper a new kind of degeneracy structure
which is due to the non-Hermitian behavior of the system. We have shown that
the pseudo-Hermitian Hamiltonians with real eigenvalues and even $\mathcal{PT%
}$-symmetry admit a degeneracy structure if the operator $\mathcal{PT}$
anticommutes with the metric operator $\eta $ which is necessarily
indefinite. We have also pointed out that our Hamiltonian $H$ admit
simultaneously two kinds of metric operators, namely, a positive definite
metric operator in the form of eq. (\ref{024b}) which does not give rise to
a degeneracy structure, and a second metric operator which is indefinite and
achieve a degeneracy structure. We have also shown that the Krein space
formulation of the Hilbert space is the convenient framework for the
implementation of unbroken $\mathcal{PT}$-symmetry for our system. Let us
discuss some implications and outlook of our analysis. \ \ \ \ \ \ \ \ \ 

$\bullet $\ \textbf{(1)}: It is useful to point out that the{\normalsize \ }$%
\mathcal{PT}${\normalsize -}symmetric four-level Hamiltonian $H$\ given in (%
\ref{H1}) represents a new class of{\normalsize \ }$\mathcal{PT}$%
{\normalsize -}symmetric Hamiltonians written in split-quaternionic form
which represent the most general traceless Hamiltonian matrix in the case of
even{\normalsize \ }$\mathcal{PT}$-symmetry by generalizing the time
reversal operator to include a matrix multiplying the complex conjugation
operator. The study of further features of this Hamiltonian will be
interesting. For instance the construction of the $\mathcal{CPT}$ inner
product in the form of Mostafazadeh $\eta _{+}$ inner product {\normalsize 
\cite{Mostafa2003}}, with $\eta _{+}\equiv $ $\mathcal{PC}$. \ 

$\bullet $\ \textbf{(2)}: The generalization to even{\normalsize \ }$%
\mathcal{PT}$-symmetry of the{\normalsize \ }$\mathcal{PT}${\normalsize - }%
and{\normalsize \ }$\mathcal{CPT}$-symmetric representations of fermionic
algebras developed by Bender {\normalsize \cite{Bender2011}} and one of the
actual authors {\normalsize \cite{Cherbal2012} }in the odd{\normalsize \ }$%
\mathcal{PT}$-symmetric case, will be also a part of our future
investigations. \ \ \ \ \ \ \ \ \ \ \ 

$\bullet $\ \textbf{(3)}: It is useful to point out that the \textit{graded
time-reversal operator in fermion Fock space} is also an interersting
outlook. In this case, the Hilbert space is neither even nor odd with
respect to time-reversal symmetry but rather has a graded structure with
regard to time reversal. The action of time reversal operator $\mathcal{T}$
may as usual be represented by $\mathcal{T}\varphi =L\varphi ^{\ast },$
where the unitary operator has however a block diagonal structure%
\begin{equation}
L=\left( 
\begin{array}{cc}
L_{+} & 0 \\ 
0 & L_{-}%
\end{array}%
\right) \,,
\end{equation}%
with $L_{+}^{2}=1,$ $L_{-}^{2}=-1.$ Thus time-reversal is even in the
bosonic subspace; it is odd in the fermionic subspace. It is a general
feature of the \textit{graded} \textit{Fock space of fermions} that
decomposes into two subspaces which are even and odd with respect to
time-reversal. Thus it will be interesting to extend our analysis and all
the works in the literature which deal with time-reversal symmetry to the
case of time-reversal symmetry in Fock space of fermions with a graded form
which can be a subject of interest for\textit{\ }$\mathcal{PT}$\textit{\ }%
community\textit{. \ \ \ \ \ \ \ \ \ \ \ \ \ \ \ \ \ \ \ \ }

{\normalsize \ }

\end{document}